\documentclass{aastex}%
\usepackage{amssymb}%
\usepackage{amsmath}%
\usepackage{amsfonts}%
\usepackage{graphicx}
\begin{document}

\begin{center}
{\LARGE Multicomponent theory of buoyancy instabilities in magnetized plasmas:
The case of magnetic field parallel to gravity}

\bigskip

{\large A. K. Nekrasov}

Institute of Physics of the Earth, Russian Academy of Sciences, 123995 Moscow, Russia

anekrasov@ifz.ru

\bigskip

{\large and}

\bigskip

{\large Mohsen Shadmehri}

Department of Physics, Golestan University, Basij Square, Gorgan, Iran

m.shadmehri@gu.ac.ir

\bigskip

{\large ABSTRACT}
\end{center}

We investigate electromagnetic buoyancy instabilities of the electron-ion
plasma with the heat flux based on not the magnetohydrodynamic (MHD)
equations, but using the multicomponent plasma approach when the momentum
equations are solved for each species. We consider a geometry in which the
background magnetic field, gravity, and stratification are directed along one
axis. The nonzero background electron thermal flux is taken into account.
Collisions between electrons and ions are included in the momentum equations.
No simplifications usual for the one-fluid MHD-approach in studying these
instabilities are used. We derive a simple dispersion relation, which shows
that the thermal flux perturbation generally stabilizes an instability for the
geometry under consideration. This result contradicts to conclusion obtained
in the MHD-approach. We show that the reason of this contradiction is the
simplified assumptions used in the MHD analysis of buoyancy instabilities and
the role of the longitudinal electric field perturbation which is not captured
by the ideal MHD equations. Our dispersion relation also shows that the medium
with the electron thermal flux can be unstable, if the temperature gradients
of ions and electrons have the opposite signs. The results obtained can be
applied to the weakly collisional magnetized plasma objects in laboratory and astrophysics.


\section{INTRODUCTION}

Thermal effects resulting in instabilities, transport, heating, structures
forming, and so on play an important role in dynamics of different plasma
objects in laboratory, space, and astrophysics. For example, the
ion-temperature-gradient-driven modes (Kadomtsev and Pogutse 1965) are used
for explaining anomalous transport in tokamak plasma experiments (Dimits et
al. 2000; Garbet 2001). Thermal conductivity influences on the Rayleigh-Taylor
instability in inertial fusion (Betti et al. 1998; Canaud et al. 2004; Lindl
et al. 2004), on the surface of the Sun (Isobe et al. 2005), in supernova
(Fryxell et al. 1991), and other astrophysical objects. Thermally stratified
fluids can be buoyantly unstable in the gravitational field. In astrophysics,
this process may, for example, operates in the stellar interiors
(Schwarzschild 1958), accretion disks (Balbus 2000, 2001), neutron stars
(Chang and Quataert 2010), hot accretion flows (Narayan et al. 2000, 2002),
galaxy clusters, and intracluster medium (ICM) (Quataert 2008; Parrish et al.
2009; Ren et al. 2009, 2010a). Analogous instabilities also exist in the
neutral atmosphere of the Earth and ocean (Gossard and Hooke 1975; Pedlosky
1982). Diversity of environments in which buoyancy (or convective)
instabilities may have the significant role, leading to turbulence and
anomalous energy and matter transport, makes these instabilities an important
object for analytical and numerical explorations.

The crucial role of convection in the transport of energy, for example, in
stellar interiors is a well-known physical process (Schwarzschild 1958).
However, theoretical efforts to understand convective energy transport in the
dilute and hot plasmas such as galaxies clusters and ICM (Sarazin 1988) have
lead to some results over recent years. As it is known, majority of the mass
of a cluster of galaxies is in the dark matter. However, around 1/6 of its
mass consists of a hot, magnetized, and low density plasma known as ICM. The
electron number density is $n_{e}\simeq10^{-3}$ to $10^{-1}$ cm$^{-3}$ at the
central parts of ICM. The electron temperature $T_{e}$ is measured of the
order of $1-15$ keV, though the ion temperature $T_{i}$ has not yet been
measured directly (Fabian et al. 2006; Sanders et al. 2010). The magnetic
field in ICM is estimated to be in the range $0.1-10$ $\mu$G depending on
where the measurement is made (Carilli and Taylor 2002). This implies a
dynamically weak magnetic field with $\beta=8\pi n_{e}T_{e}/B^{2}%
\approx200-2000$. Under conditions given above, the Larmor radius of electrons
and ions ($T_{i}\sim T_{e}$) is many orders of magnitude smaller than the mean
free path. Thus, the ICM is classified as a weakly collisional plasma (Carilli
and Taylor 2002) possessing anisotropic transport due to the magnetic field.

In recent past, for such plasmas in the framework of the ideal MHD
supplemented by an anisotropic heat flux along the magnetic field, there were
found some new convective instabilities for the case when a heat flux plays
the significant role (Balbus 2000, 2001; Quataert 2008; Ren et al. 2009,
2010a, 2010b). One of these instabilities, at the absence of the background
thermal flux, has been shown to arise when the temperature increases in the
direction of gravity which is perpendicular to the background magnetic field.
This is so-called the magnetothermal instability (MTI) (Balbus 2000, 2001).
The other instability, the heat buoyancy instability (HBI) (Quataert 2008),
has been found to arise at the presence of the background heat flux when the
temperature decreases along gravity parallel to the magnetic field.
Anisotropic dissipative effects have been included by Ren et al. (2010a, 2010b).

Theoretical models applied for study of buoyancy instabilities in
astrophysical objects with a heat flux are based on the one-fluid ideal
(Balbus 2000, 2001; Quataert 2008; Ren et al. 2009; Chang and Quataert 2010)
and nonideal (Ren et al. 2010a, 2010b) MHD equations. By using of these
equations one can comparatively easily to consider any problems. However, the
ideal MHD does not capture some important effects which can be taken into
account by using a multi-fluid plasma approach. One of such effects is the
nonzero longitudinal electric field perturbation along the background magnetic
field. An importance of involving this component due to multi-fluid effects
and shortcomings of the ideal MHD were emphasized, for example, for the
acceleration of solar flare electrons by inertial Alfv\'{e}n waves (McClements
and Fletcher 2009), at consideration of structures of electromagnetic fields
and plasma flows in pulsar magnetosphere (Kojima and Oogi 2009), for the
acceleration of relativistic ions, electrons, and positrons in shock waves
(Takahashi et al. 2009), and in a gyrofluid description of Alfv\'{e}nic
turbulence (Bian and Kontar 2010). As we show here, the contribution of
currents due to this (small in the present case) parallel electric field to
the dispersion relation can be of the same order of magnitude as that due to
transverse electric field components. Besides, the MHD equations do not take
into account the existence of various charged and neutral species with
different masses and electric charges and their collisions between each others
and therefore can not be applied to multicomponent systems. In some cases, the
standard methods used in the MHD lead to conclusions that are different from
those obtained by the method using the electric field perturbations (the
$\mathbf{E}$-approach). One such an example concerning the contribution of the
electron-ion collisions to the dispersion relation for the MHD waves in the
two-component magnetized plasma was considered by Nekrasov (2009c). A
multicomponent approach has been used in (Nekrasov 2008, 2009a, 2009b, 2009c,
2009d), where the streaming instabilities of rotating astrophysical objects
(accretion disks, molecular clouds and so on) have been investigated.

A study of buoyancy instabilities with the electron heat flux by the
multicomponent $\mathbf{E}$-approach has been performed by Nekrasov and
Shadmehri (2010). The geometry has been considered in which the gravity is
perpendicular to the background magnetic field and the background heat flux is
absent. Solution of the dispersion relation obtained in this paper differs
from solution of the same problem found from the ideal MHD (Balbus 2000).

In this paper, we apply a multicomponent approach to study buoyancy
instabilities in magnetized electron-ion plasmas with the background electron
thermal flux. We consider the geometry in which the gravity, stratification,
background magnetic field and thermal flux are all directed along one ($z$-)
axis. For generality, we include collisions between electrons and ions in the
momentum equations. At the consideration of the perturbed heat flux, we adopt
that cyclotron frequencies of species are much larger than their collision
frequencies. Such conditions are typical for many laboratory, space, and
astrophysical plasmas. In this case, the heat flux is anisotropic and directed
along the magnetic field lines (Braginskii 1965). However in other respects,
the relation between the cyclotron and collision frequencies is arbitrary in
the general expressions for the perturbed values. We derive the dispersion
relation for cases, in which the background heat flux is present or absent.
This gives a possibility to compare these two cases. Solutions of the
dispersion relation are discussed.

The paper is organized as follows. In Sect. 2, the fundamental equations are
given. An equilibrium state is considered in Sect. 3. Perturbed ion velocity,
number density, and thermal pressure are obtained in Sect. 4. In Sect. 5, we
consider the perturbed velocity and temperature for electrons. Components of
the dielectric permeability tensor are found in Sect. 6. Dispersion relation
is derived and considered in the collisionless and collisional cases in Sect.
7. Discussion of the results obtained and comparison with the MHD results are
provided in Sect. 8. In Sect. 9, we give conclusive remarks.

\bigskip

\section{BASIC EQUATIONS}

We start with the following equations for ions:
\begin{equation}
\frac{\partial\mathbf{v}_{i}}{\partial t}\mathbf{=-}\frac{\mathbf{\nabla}%
p_{i}}{m_{i}n_{i}}+\mathbf{g+}\frac{q_{i}}{m_{i}}\mathbf{E}+\frac{q_{i}}%
{m_{i}c}\mathbf{v}_{i}\times\mathbf{B}-\nu_{ie}\left(  \mathbf{v}%
_{i}-\mathbf{v}_{e}\right)  ,
\end{equation}
the momentum equation,
\begin{equation}
\frac{\partial n_{i}}{\partial t}+\mathbf{\nabla}\cdot n_{i}\mathbf{v}_{i}=0,
\end{equation}
the continuity equation, and
\begin{equation}
\frac{\partial p_{i}}{\partial t}+\mathbf{v}_{i}\cdot\mathbf{\nabla}%
p_{i}+\gamma p_{i}\mathbf{\nabla}\cdot\mathbf{v}_{i}=0,
\end{equation}
the pressure equation. The corresponding equations for electrons are:
\begin{equation}
\mathbf{0=-}\frac{\mathbf{\nabla}p_{e}}{n_{e}}+q_{e}\mathbf{E}+\frac{q_{e}}%
{c}\mathbf{v}_{e}\times\mathbf{B}-m_{e}\nu_{ei}\left(  \mathbf{v}%
_{e}-\mathbf{v}_{i}\right)  ,
\end{equation}%
\begin{equation}
\frac{\partial n_{e}}{\partial t}+\mathbf{\nabla}\cdot n_{e}\mathbf{v}_{e}=0,
\end{equation}%
\begin{equation}
\frac{\partial p_{e}}{\partial t}+\mathbf{v}_{e}\cdot\mathbf{\nabla}%
p_{e}+\gamma p_{e}\mathbf{\nabla}\cdot\mathbf{v}_{e}=\lambda-\left(
\gamma-1\right)  \mathbf{\nabla\cdot q}_{e},
\end{equation}%
\begin{equation}
\frac{\partial T_{e}}{\partial t}+\mathbf{v}_{e}\cdot\mathbf{\nabla}%
T_{e}+\left(  \gamma-1\right)  T_{e}\mathbf{\nabla}\cdot\mathbf{v}_{e}%
=\frac{\lambda}{n_{e}}-\left(  \gamma-1\right)  \frac{1}{n_{e}}\mathbf{\nabla
\cdot q}_{e},
\end{equation}
the temperature equation, where $\mathbf{q}_{e}$ is the electron heat flux
(Braginskii 1965). We neglect inertia of the electrons. In (1)-(7), $q_{j}$
and $m_{j}$ are the charge and mass of species $j=i,e$, $\mathbf{v}_{j}$ is
the hydrodynamic velocity, $n_{j}$ is the number density, $p_{j}=n_{j}T_{j}$
is the thermal pressure, $T_{j}$ is the temperature, $\nu_{ie}$ ($\nu_{ei}$)
is the collision frequency of ions (electrons) with electrons (ions),
$\mathbf{g}$ is gravity, $\mathbf{E}$\textbf{\ }and $\mathbf{B}$ are the
electric and magnetic fields, $c$ is the speed of light in vacuum, and
$\gamma$ is the adiabatic constant. At the consideration of the electron heat
flux, we will assume the electrons to be magnetized when their cyclotron
frequency $\omega_{ce}=q_{e}B/m_{e}c\gg\nu_{ee}(\nu_{ei})$, where $\nu
_{ee}(\nu_{ei})$ is the electron-electron (ion) collision frequency. In this
case, the electron thermal flux is mainly directed along the magnetic field,%
\begin{equation}
\mathbf{q}_{e}=-\chi_{e}\mathbf{b}\left(  \mathbf{b\cdot\nabla}\right)  T_{e},
\end{equation}
where $\chi_{e}$ is the electron thermal conductivity coefficient and
$\mathbf{b=B/}B$ is the unit vector along the magnetic field (Braginskii
1965). However in the momentum equations (1) and (4), we keep for generality
an arbitrary relation between $\omega_{ci}(\omega_{ce})$ and $\nu_{ie}%
(\nu_{ei})$ ($\omega_{ci}=q_{i}B/m_{i}c$ and $\nu_{ie}$ are the ion cyclotron
and ion-electron collision frequencies, respectively), having in mind that
some expressions obtained below can be applied for collisional objects. The
term $\lambda$ compensates the temperature change as a result of the
equilibrium heat flux. We take only into account the electron thermal
conductivity by (8), because the corresponding ion conductivity is
considerably smaller (Braginskii 1965). For generality, we assume the electron
and ion temperatures to be different. However, we do not involve, for
simplicity, the terms describing the energy exchange between ions and
electrons in (3), (6), and (7). Thus, our treatment is available for cases in
which such an exchange is not effective or when there is a strong temperature
coupling of species. In the last case, one can set $T_{i}\simeq T_{e}$. This
issue is considered in more detail in Sect. 8.

Electromagnetic equations are Faraday's law
\begin{equation}
\mathbf{\nabla\times E=-}\frac{1}{c}\frac{\partial\mathbf{B}}{\partial t}%
\end{equation}
and Ampere`s law
\begin{equation}
\mathbf{\nabla\times B=}\frac{4\pi}{c}\mathbf{j,}%
\end{equation}
where $\mathbf{j=}\sum_{j}q_{j}n_{j}\mathbf{v}_{j}.$ We consider the wave
processes with typical timescales much larger than the time the light spends
to cover the wavelength of perturbations. In this case, one can neglect the
displacement current in (10) that results in quasineutrality both in
electromagnetic and purely electrostatic perturbations. The magnetic field
$\mathbf{B}$ includes the background magnetic field $\mathbf{B}_{0}$, the
magnetic field $\mathbf{B}_{0cur}$ of the background current (when it
presents), and the perturbed magnetic field.

\bigskip

\section{EQUILIBRIUM STATE}

At first, we consider an equilibrium state. We assume that background
velocities are absent. In this paper, we study configuration in which the
background magnetic field, gravity, and stratification are directed along the
$z$-axis. Let, for definiteness, $\mathbf{g}$ be $\mathbf{g=-z}g$, where $g>0$
and $\mathbf{z}$ is the unit vector along the $z$-direction. Then, (1) and (4)
give
\begin{equation}
g_{i}=-\frac{1}{m_{i}n_{i0}}\frac{\partial p_{i0}}{\partial z}=g-\frac{q_{i}%
}{m_{i}}E_{0},
\end{equation}%
\begin{equation}
g_{e}=-\frac{1}{m_{i}n_{e0}}\frac{\partial p_{e0}}{\partial z}=\frac{q_{i}%
}{m_{i}}E_{0},
\end{equation}
where (and below) the index $0$ denotes equilibrium values. Here and below, we
assume that $q_{i}=-q_{e}$. For convenience of notations, we do not use that
$n_{i0}=n_{e0}$ for the two-component plasma up to a point where it will be
necessary. We see that equilibrium distributions of ions and electrons
influence on each other through the background electric field $E_{0}$. In the
case $n_{i0}=n_{e0}$ and $T_{i0}=T_{e0}$, we have $g_{i}=g_{e}=$ $g/2$. Thus,
we obtain $E_{0}=m_{i}g/2q_{i}$. The presence of the third component, for
example, of the cold dust grains with the charge $q_{d}$ and mass $m_{d}\gg
m_{i}$ results in other value of $E_{0}=m_{d}g/q_{d}$. In this case, the ions
and electrons are in equilibrium under the action of the thermal pressure and
equilibrium electric field, being $g_{i}\simeq-g_{e}$, if $q_{i}m_{d}\gg
q_{d}m_{i}$.

\bigskip

\section{LINEAR\ ION\ PERTURBATIONS}

Let us write (1)-(3) for ions in the linear approximation,%
\begin{equation}
\frac{\partial\mathbf{v}_{i1}}{\partial t}\mathbf{=-}\frac{\mathbf{\nabla
}p_{i1}}{m_{i}n_{i0}}+\frac{\mathbf{\nabla}p_{i0}}{m_{i}n_{i0}}\frac{n_{i1}%
}{n_{i0}}+\mathbf{F}_{i1}+\frac{q_{i}}{m_{i}c}\mathbf{v}_{i1}\times
\mathbf{B}_{0},
\end{equation}%
\begin{equation}
\frac{\partial n_{i1}}{\partial t}+v_{i1z}\frac{\partial n_{i0}}{\partial
z}+n_{i0}\mathbf{\nabla}\cdot\mathbf{v}_{i1}=0,
\end{equation}%
\begin{equation}
\frac{\partial p_{i1}}{\partial t}+v_{i1z}\frac{\partial p_{i0}}{\partial
z}+\gamma p_{i0}\mathbf{\nabla}\cdot\mathbf{v}_{i1}=0,
\end{equation}
where
\begin{equation}
\mathbf{F}_{i1}=\frac{q_{i}}{m_{i}}\mathbf{E}_{1}-\nu_{ie}\left(
\mathbf{v}_{i1}-\mathbf{v}_{e1}\right)  ,
\end{equation}
and the index $1$ denotes the perturbed variables. Below, we solve these
equations to find the perturbed velocity of ions in an inhomogeneous medium.

\subsection{Perturbed velocity of ions}

Applying the operator $\partial/\partial t$ to (13) and using (14) and (15),
we obtain equation containing only the ion velocity%
\begin{equation}
\frac{\partial^{2}\mathbf{v}_{i1}}{\partial t^{2}}\mathbf{=-}g_{i}%
\mathbf{\nabla}v_{i1z}+\frac{1}{m_{i}n_{i0}}\left[  \left(  \gamma-1\right)
\left(  \mathbf{\nabla}p_{i0}\right)  +\gamma p_{i0}\mathbf{\nabla}\right]
\mathbf{\nabla}\cdot\mathbf{v}_{i1}+\frac{\partial\mathbf{F}_{i1}}{\partial
t}+\frac{q_{i}}{m_{i}c}\frac{\partial\mathbf{v}_{i1}}{\partial t}%
\times\mathbf{B}_{0}.
\end{equation}
From this equation, we can find solutions for the components of $\mathbf{v}%
_{i1}$. For simplicity, we assume that $\partial/\partial x=0$ because a
system is symmetric in the transverse direction relative to the $z$-axis. The
$x$-component of (17) has the simple form
\begin{equation}
\frac{\partial v_{i1x}}{\partial t}\mathbf{=}F_{i1x}+\omega_{ci}v_{i1y}.
\end{equation}
Here $\omega_{ci}=q_{i}B_{0}/m_{i}c$. For the $y$-component of (17), we
obtain:%
\begin{equation}
\frac{\partial^{2}v_{i1y}}{\partial t^{2}}\mathbf{=-}g_{i}\frac{\partial
v_{i1z}}{\partial y}+c_{si}^{2}\frac{\partial}{\partial y}\mathbf{\nabla}%
\cdot\mathbf{v}_{i1}+\frac{\partial F_{i1y}}{\partial t}-\omega_{ci}%
\frac{\partial v_{i1x}}{\partial t},
\end{equation}
where, $c_{si}=\left(  \gamma T_{i0}/m_{i}\right)  ^{1/2}$ is the ion sound
velocity. Using (18), equation for $v_{i1y}$ is given from (19) as follows
\begin{equation}
\left(  \frac{\partial^{2}}{\partial t^{2}}+\omega_{ci}^{2}\right)
v_{i1y}-Q_{i1y}\mathbf{=}\frac{\partial P_{i1}}{\partial y}.
\end{equation}
Then from (18), we obtain%
\begin{equation}
\frac{\partial}{\omega_{ci}\partial t}\left[  \left(  \frac{\partial^{2}%
}{\partial t^{2}}+\omega_{ci}^{2}\right)  v_{i1x}-Q_{i1x}\right]
\mathbf{=}\frac{\partial P_{i1}}{\partial y}.
\end{equation}
In (20) and (21), we have introduced the following notations:%
\begin{equation}
P_{i1}=\mathbf{-}g_{i}v_{i1z}+c_{si}^{2}\mathbf{\nabla}\cdot\mathbf{v}_{i1},
\end{equation}%
\begin{equation}
Q_{i1x}=\omega_{ci}F_{i1y}+\frac{\partial F_{i1x}}{\partial t},
\end{equation}%
\begin{equation}
Q_{i1y}=-\omega_{ci}F_{i1x}+\frac{\partial F_{i1y}}{\partial t}.
\end{equation}
The value $P_{i1}$ defines the ion pressure perturbation (see 15). We see from
(20) and (21) that the thermal pressure effect on the velocity $v_{i1x}$ is
much larger than that on $v_{i1y}$ when $\partial/\partial t\ll\omega_{ci}$.

The $z$-component of (17) takes the form
\begin{equation}
\frac{\partial}{\partial t}\left(  \frac{\partial v_{i1z}}{\partial t}%
-F_{i1z}\right)  \mathbf{=-}g_{i}\frac{\partial v_{i1z}}{\partial z}+\left[
\left(  1-\gamma\right)  g_{i}+c_{si}^{2}\frac{\partial}{\partial z}\right]
\mathbf{\nabla}\cdot\mathbf{v}_{i1}.
\end{equation}
To obtain equation only for $v_{i1z}$, we need to express $\mathbf{\nabla
}\cdot\mathbf{v}_{i1}$ through $v_{i1z}$. Differentiating (20) with respect to
$y$ and using expression (22), we find
\begin{equation}
L_{1}\mathbf{\nabla}\cdot\mathbf{v}_{i1}\mathbf{=}L_{2}v_{i1z}+\frac{\partial
Q_{i1y}}{\partial y},
\end{equation}
where the following operators are introduced:%
\begin{equation}
L_{1}=\frac{\partial^{2}}{\partial t^{2}}+\omega_{ci}^{2}\mathbf{-}c_{si}%
^{2}\frac{\partial^{2}}{\partial y^{2}},
\end{equation}%
\begin{equation}
L_{2}=\left(  \frac{\partial^{2}}{\partial t^{2}}+\omega_{ci}^{2}\right)
\frac{\partial}{\partial z}-g_{i}\frac{\partial^{2}}{\partial y^{2}}.
\end{equation}
We can derive equation for the longitudinal velocity $v_{i1z}$, substituting
$\mathbf{\nabla}\cdot\mathbf{v}_{i1}$ found from (26) into (25),%
\begin{equation}
L_{3}v_{i1z}\mathbf{=}L_{1}\frac{\partial F_{i1z}}{\partial t}+L_{4}%
\frac{\partial Q_{i1y}}{\partial y},
\end{equation}
where operators $L_{3}$ and $L_{4}$ have the form
\begin{align}
L_{3}  & =\left(  \frac{\partial^{2}}{\partial t^{2}}+\omega_{ci}^{2}\right)
\frac{\partial^{2}}{\partial t^{2}}-c_{si}^{2}\left(  \frac{\partial^{2}%
}{\partial y^{2}}+\frac{\partial^{2}}{\partial z^{2}}\right)  \frac
{\partial^{2}}{\partial t^{2}}-c_{si}^{2}\omega_{ci}^{2}\frac{\partial^{2}%
}{\partial z^{2}}\\
& +\gamma g_{i}\left(  \frac{\partial^{2}}{\partial t^{2}}+\omega_{ci}%
^{2}\right)  \frac{\partial}{\partial z}+c_{si}^{2}\frac{\partial L_{1}}%
{L_{1}\partial z}L_{2}+\left(  1-\gamma\right)  g_{i}^{2}\frac{\partial^{2}%
}{\partial y^{2}},\nonumber
\end{align}

\begin{equation}
L_{4}=\left(  1-\gamma\right)  g_{i}+c_{si}^{2}\left(  \frac{\partial
}{\partial z}-\frac{\partial L_{1}}{L_{1}\partial z}\right)  .
\end{equation}
For obtaining expression (30), we have used expressions (27) and (28).

It is easy to see that at the absence of the background magnetic field and
without taking into account electromagnetic perturbations and collisions with
electrons (the right hand-side of 29), equation $L_{3}v_{i1z}=0$ describes the
ion sound and internal gravity waves. In this case, a sum of the last two
terms on the right hand-side of expression (30) is equal to $-c_{si}^{2}%
\omega_{bi}^{2}\frac{\partial^{2}}{\partial y^{2}}$, where $\omega_{bi}$ is
the (ion) Brunt-V\"{a}is\"{a}l\"{a} frequency equal to%
\begin{equation}
\omega_{bi}^{2}=\frac{g_{i}}{c_{si}^{2}}\left[  \left(  \gamma-1\right)
g_{i}+\frac{\partial c_{si}^{2}}{\partial z}\right]  .
\end{equation}
Thus, we have obtained a result corresponding to perturbations in the neutral
atmosphere (Gossard and Hooke 1975). However, we see that the existence of the
background magnetic field considerably modifies the operator $L_{3}$. We note
that the right hand-side of (29) describes a connection between ions and
electrons through the electric field $\mathbf{E}_{1}$ and collisions.

\subsection{Specific case for ions}

So far, we did not make any simplifications and all the equations and
expressions obtained above are given in their general form. This permits us to
investigate different limiting cases. Further, we consider perturbations with
a frequency much lower than the ion cyclotron frequency and the transverse
wavelengths much larger than the ion Larmor radius. Such perturbations are of
interest for both laboratory and astrophysical plasmas. Besides, we
investigate a part of the frequency spectrum in the region lower than the ion
sound frequency. Thus, we set%
\begin{equation}
\omega_{ci}^{2}\gg\frac{\partial^{2}}{\partial t^{2}},c_{si}^{2}\frac
{\partial^{2}}{\partial y^{2}};c_{si}^{2}\frac{\partial^{2}}{\partial z^{2}%
}\gg\frac{\partial^{2}}{\partial t^{2}}.
\end{equation}
In this case, operators (27), (28), (30), and (31) take the form%
\begin{align}
L_{1}  & \simeq\omega_{ci}^{2},L_{2}\simeq\omega_{ci}^{2}\frac{\partial
}{\partial z},\\
L_{3}  & =-\omega_{ci}^{2}\left[  \left(  c_{si}^{2}\frac{\partial}{\partial
z}-\gamma g_{i}\right)  \frac{\partial}{\partial z}-\frac{\partial^{2}%
}{\partial t^{2}}\right]  ,\nonumber\\
L_{4}  & =\left(  1-\gamma\right)  g_{i}+c_{si}^{2}\frac{\partial}{\partial
z}.\nonumber
\end{align}
Also, an additional condition
\begin{equation}
\omega_{ci}^{2}\frac{\partial^{2}}{\partial t^{2}}\gg c_{si}^{2}\frac{\partial
c_{si}^{2}}{\partial z}\frac{\partial^{3}}{\partial y^{2}\partial z}%
\end{equation}
must be satisfied for operator $L_{3}$ to have a given form (34). The small
corrections in operators $L_{3}$ and $L_{4}$ are needed to be kept because
some main terms in expressions for the ion and electron velocities are equal
to each other (see below). Therefore, when calculating the electric current,
these main terms will be canceled and small corrections to velocities will
only contribute to the current. The applicability of condition (35) and other
conditions used below will be discussed in Sect. 8.

For cases represented by inequalities (33) and (35) when the operators $L_{i}
$, $i=1,2,3,4,$ have the form (34), equations for $v_{i1z}$ and
$\mathbf{\nabla}\cdot\mathbf{v}_{i1}$ become%
\begin{equation}
\left[  \left(  c_{si}^{2}\frac{\partial}{\partial z}-\gamma g_{i}\right)
\frac{\partial}{\partial z}-\frac{\partial^{2}}{\partial t^{2}}\right]
v_{i1z}\mathbf{=-}\frac{\partial F_{i1z}}{\partial t}-\left[  \left(
1-\gamma\right)  g_{i}+c_{si}^{2}\frac{\partial}{\partial z}\right]
\frac{\partial Q_{i1y}}{\omega_{ci}^{2}\partial y},
\end{equation}%
\begin{equation}
\mathbf{\nabla}\cdot\mathbf{v}_{i1}\simeq\frac{\partial v_{i1z}}{\partial
z}+\frac{\partial Q_{i1y}}{\omega_{ci}^{2}\partial y}.
\end{equation}

\subsection{Ion perturbations in the Fourier transformation}

Calculations show that some main terms in expressions for $v_{i1z}$ (when
calculating the current), $\mathbf{\nabla}\cdot\mathbf{v}_{i1}$ and $P_{i1} $
are canceled. Therefore, the small terms proportional to inhomogeneity must be
taken into account. To do this correctly, we can not apply the Fourier
transformation to (36) and (37) to find the variable $P_{i1}$. However,
firstly, we should apply the operator $\partial/\partial z$ to this variable
for using (36). It is analogous to obtaining the term $\partial c_{s}%
^{2}/\partial z$ in expression (32) for the Brunt-V\"{a}is\"{a}l\"{a}
frequency. After that, we can apply the Fourier transformation in a local
approximation assuming the linear perturbations to be proportional to
$\exp(i\mathbf{kr-}i\omega t)$. As a result, we obtain for the
Fourier-components $v_{i1zk}$, $\mathbf{k}\cdot\mathbf{v}_{i1k}$, and
$P_{i1k}$, where $k=\left(  \mathbf{k,}\omega\right)  $, the following
expressions:
\begin{equation}
v_{i1zk}=-i\frac{\omega}{k_{z}^{2}c_{si}^{2}}\left(  1-i\frac{\gamma g_{i}%
}{k_{z}c_{si}^{2}}\right)  F_{i1zk}-\frac{k_{y}}{k_{z}\omega_{ci}^{2}}\left(
1-i\frac{g_{i}}{k_{z}c_{si}^{2}}\right)  Q_{i1yk},
\end{equation}%
\begin{equation}
\mathbf{k}\cdot\mathbf{v}_{i1k}\mathbf{=-}i\frac{\omega}{k_{z}c_{si}^{2}%
}\left(  1-i\frac{\gamma g_{i}}{k_{z}c_{si}^{2}}\right)  F_{i1zk}+i\frac
{k_{y}}{k_{z}}\frac{g_{i}}{c_{si}^{2}\omega_{ci}^{2}}Q_{i1yk},
\end{equation}%
\begin{align}
P_{i1k}  & =\frac{\omega}{k_{z}}F_{i1zk}-i\frac{\omega}{k_{z}^{2}c_{si}^{2}%
}\left[  \left(  \gamma-1\right)  g_{i}+\frac{\partial c_{si}^{2}}{\partial
z}\right]  F_{i1zk}\\
& +i\frac{k_{y}g_{i}}{k_{z}^{2}c_{si}^{2}\omega_{ci}^{2}}\left[  \left(
\gamma-1\right)  g_{i}+\frac{\partial c_{si}^{2}}{\partial z}-\omega^{2}%
\frac{c_{si}^{2}}{g_{i}}\right]  Q_{i1yk},\nonumber
\end{align}
where $g_{i}/k_{z}c_{si}^{2}\ll1$. In expressions (38) and (39), we have
omitted additional small terms at $Q_{i1yk}$ which are needed for calculation
of $P_{i1k}$.

When calculating the current along the $z$-axis, the main term $\sim Q_{i1yk}
$ in (38) will be canceled with the corresponding electron term. The
contribution of the first term $\sim F_{i1zk}$ to this current has, as we will
see below, the same order of magnitude for the buoyancy instabilities as
contribution of the term $\sim g_{i}Q_{i1yk}$, i.e. $F_{i1zk}\sim\left(
k_{y}g_{i}/\omega\omega_{ci}^{2}\right)  Q_{i1yk}$. The same relates to
expressions (39) and (40). Thus, the longitudinal electric field perturbation
$E_{1z}$ containing in $F_{i1z}$ must be taken into account. However, in the
ideal MHD, this field is absent. We see from expressions (38) and (39) that
$\mathbf{\nabla}\cdot\mathbf{v}_{i1}\sim(g_{i}/c_{si}^{2})v_{i1z}$. This
relation is the same as that for internal gravity waves in the Earth's
atmosphere (see, e.g., Nekrasov 1994). Using expression (40), we can find
velocities $v_{i1yk}$ and $v_{i1xk}$ from (20) and (21), correspondingly.

\subsection{Perturbed ion number density and pressure}

It is followed from above that $\mathbf{\nabla}\cdot\mathbf{v}_{i1}\sim
v_{i1z}/H$, where $H$ is the inhomogeneity scale height ($H\sim c_{si}%
^{2}/g_{i}$). Thus, the last two terms in (14) and (15) are of the same order
of magnitude. Let us find the perturbed ion number density and pressure in the
Fourier-representation. From (14), (38) and (39), we obtain%
\begin{equation}
\frac{n_{i1k}}{n_{i0}}=-i\frac{1}{k_{z}c_{si}^{2}}F_{i1zk}-i\frac{k_{y}}%
{k_{z}c_{si}^{2}\omega\omega_{ci}^{2}}\left[  \left(  \gamma-1\right)
g_{i}+\frac{\partial c_{si}^{2}}{\partial z}\right]  Q_{i1yk}.
\end{equation}
Equation (15) gives $\partial p_{i1}/\partial t=-m_{i}n_{i0}P_{i1}$. Thus, we
have, using (40),%
\begin{equation}
\frac{p_{i1k}}{p_{i0}}=-i\frac{\gamma}{k_{z}c_{si}^{2}}F_{i1zk}+\frac{\gamma
k_{y}g_{i}}{k_{z}^{2}c_{si}^{4}\omega\omega_{ci}^{2}}\left[  \left(
\gamma-1\right)  g_{i}+\frac{\partial c_{si}^{2}}{\partial z}-\omega^{2}%
\frac{c_{si}^{2}}{g_{i}}\right]  Q_{i1yk}.
\end{equation}

Comparing (41) and (42), we see that the relative perturbation of the pressure
due to the transverse electric force $Q_{i1yk}$ is much smaller than the
relative perturbation of the number density. However, these relative
perturbations as a result of the action of the longitudinal electric force
$F_{i1zk}$ have the same order of magnitude (see Sect. 4.3). Thus,
$p_{i1k}/p_{i0}$ $\sim n_{i1k}/n_{i0}$. This result contradicts a supposition
$p_{i1k}/p_{i0}$ $\ll n_{i1k}/n_{i0}$ adopted in the MHD analysis of buoyancy
instabilities (Balbus 2000, 2001; Quataert 2008). From results obtained below,
it is followed that, as we already have noted above, the both terms on the
right hand-side of (41) have the same order of magnitude.

\bigskip\

\section{LINEAR\ ELECTRON\ PERTURBATIONS}

Equations for the electrons in the linear approximation are the following:%
\begin{equation}
\mathbf{0=-}\frac{\mathbf{\nabla}p_{e1}}{n_{e0}}+\frac{\mathbf{\nabla}p_{e0}%
}{n_{e0}}\frac{n_{e1}}{n_{e0}}+\mathbf{F}_{e1}+\frac{q_{e}}{c}\mathbf{v}%
_{e1}\times\mathbf{B}_{0},
\end{equation}%
\begin{equation}
\frac{\partial n_{e1}}{\partial t}+v_{e1z}\frac{\partial n_{e0}}{\partial
z}+n_{e0}\mathbf{\nabla}\cdot\mathbf{v}_{e1}=0,
\end{equation}%
\begin{equation}
\frac{\partial p_{e1}}{\partial t}+v_{e1z}\frac{\partial p_{e0}}{\partial
z}+\gamma p_{e0}\mathbf{\nabla}\cdot\mathbf{v}_{e1}=-\left(  \gamma-1\right)
\mathbf{\nabla\cdot q}_{e1},
\end{equation}%
\begin{equation}
\frac{\partial T_{e1}}{\partial t}+v_{e1z}\frac{\partial T_{e0}}{\partial
z}+\left(  \gamma-1\right)  T_{e0}\mathbf{\nabla}\cdot\mathbf{v}_{e1}=-\left(
\gamma-1\right)  \frac{1}{n_{e0}}\mathbf{\nabla\cdot q}_{e1},
\end{equation}%
\begin{equation}
\mathbf{q}_{e1}=-\mathbf{b}_{1}\chi_{e0}\frac{\partial T_{e0}}{\partial
z}-\mathbf{b}_{0}\chi_{e0}\frac{\partial T_{e1}}{\partial z}-\mathbf{b}%
_{0}\chi_{e1}\frac{\partial T_{e0}}{\partial z},
\end{equation}%
\begin{equation}
\mathbf{F}_{e1}=q_{e}\mathbf{E}_{1}-m_{e}\nu_{ei}\left(  \mathbf{v}%
_{e1}-\mathbf{v}_{i1}\right)  .
\end{equation}
In (47), $\chi_{e1}=5\chi_{e0}T_{e1}/2T_{e0}$ is the perturbation of the
thermal flux conductivity coefficient $\chi_{e}$ which is proportional to
$T_{e}^{5/2}$ (Spitzer 1962; Braginskii 1965). The perturbation of the unit
magnetic vector $\mathbf{b}_{1}$ is equal to $b_{1x,y}=B_{1x,y}/B_{0}$ and
$b_{1z}=0$. The thermal flux in equilibrium is $\mathbf{q}_{e0}=-\mathbf{b}%
_{0}\chi_{e0}\frac{\partial T_{e0}}{\partial z}$.

We have seen above at consideration of the ion perturbations that the terms
$\sim1/H^{2}$ are needed to be kept (see the last term in 40). Therefore, we
also keep such terms for the electrons.

\subsection{Equation for the electron temperature perturbation}

Let us now find equation for the electron temperature perturbation. Expression
$\mathbf{\nabla\cdot q}_{e1}$, where $\mathbf{q}_{e1}$ is defined by (47), is
given by%
\begin{equation}
\mathbf{\nabla\cdot q}_{e1}=\frac{\partial q_{e1y}}{\partial y}+\frac{\partial
q_{e1z}}{\partial z}=-\chi_{e0}\frac{\partial T_{e0}}{\partial z}\frac
{1}{B_{0}}\frac{\partial B_{1y}}{\partial y}-\chi_{e0}\frac{\partial^{2}%
T_{e1}}{\partial z^{2}}-2\frac{\partial\chi_{e0}}{\partial z}\frac{\partial
T_{e1}}{\partial z}-\frac{\partial^{2}\chi_{e0}}{\partial z^{2}}T_{e1}.
\end{equation}
Substituting this expression into (46), we obtain%
\begin{equation}
D_{1}T_{e1}=-v_{e1z}\frac{\partial T_{e0}}{\partial z}-\left(  \gamma
-1\right)  T_{e0}\mathbf{\nabla}\cdot\mathbf{v}_{e1}+\left(  \gamma-1\right)
\frac{\chi_{e0}}{n_{e0}}\frac{\partial T_{e0}}{\partial z}\frac{\partial
B_{1y}}{B_{0}\partial y},
\end{equation}
where the operator $D_{1}$ is defined by%
\begin{equation}
D_{1}=\left[  \frac{\partial}{\partial t}-\left(  \gamma-1\right)  \frac
{1}{n_{e0}}\left(  \chi_{e0}\frac{\partial^{2}}{\partial z^{2}}+2\frac
{\partial\chi_{e0}}{\partial z}\frac{\partial}{\partial z}+\frac{\partial
^{2}\chi_{e0}}{\partial z^{2}}\right)  \right]  .
\end{equation}

\subsection{Perturbed velocity and temperature of electrons}

We further find equations for components of the perturbed velocity of
electrons. The $x$-component of (43) has a simple form, i.e.
\begin{equation}
v_{e1y}=-\frac{1}{m_{e}\omega_{ce}}F_{e1x},
\end{equation}
where $\omega_{ce}=q_{e}B_{0}/m_{e}c$. Applying the operator $\partial
/\partial t$ to the $y$-component of (43) and using (45) and (49), we obtain%
\begin{align}
\frac{\partial}{\partial t}\left(  v_{e1x}-\frac{1}{m_{e}\omega_{ce}}%
F_{e1y}\right)   & \mathbf{=}\mathbf{-}\frac{1}{\omega_{ci}}\frac{\partial
P_{e1}}{\partial y}-\left(  \gamma-1\right)  \frac{\chi_{e0}}{m_{e}\omega
_{ce}n_{e0}}\frac{\partial T_{e0}}{\partial z}\frac{\partial^{2}B_{1y}}%
{B_{0}\partial y^{2}}\\
& +\frac{1}{m_{e}\omega_{ce}}\left(  D_{1}-\frac{\partial}{\partial t}\right)
\frac{\partial T_{e1}}{\partial y}.\nonumber
\end{align}
Here
\begin{equation}
P_{e1}=-g_{e}v_{e1z}+c_{se}^{2}\mathbf{\nabla}\cdot\mathbf{v}_{e1},
\end{equation}
where $c_{se}^{2}=\gamma p_{e0}/$ $m_{i}n_{e0}$. The variable $P_{e1}$ is
analogous to $P_{i1}$ (see 22) and defines the electron pressure perturbation.
But for electrons, their pressure perturbation is also affected by the thermal
conductivity (see 45). The $z$-component of (43) takes the form
\begin{equation}
0\mathbf{=-}\frac{1}{n_{e0}}\frac{\partial p_{e1}}{\partial z}+\frac{1}%
{n_{e0}}\frac{\partial p_{e0}}{\partial z}\frac{n_{e1}}{n_{e0}}+F_{e1z}.
\end{equation}
We can express $\mathbf{\nabla}\cdot\mathbf{v}_{e1}$ which is contained in
(54) through $v_{e1z}$, using (52),%
\begin{equation}
\mathbf{\nabla}\cdot\mathbf{v}_{e1}=\frac{\partial v_{e1z}}{\partial z}%
-\frac{1}{m_{e}\omega_{ce}}\frac{\partial F_{e1x}}{\partial y}.
\end{equation}
This expression for electrons is analogous to that for ions (see 37).

We further consider perturbations with the dynamical frequency $\partial
/\partial t$ satisfying the following conditions:%
\begin{equation}
\frac{\chi_{e0}}{n_{e0}}\frac{\partial^{2}}{\partial z^{2}}\gg\frac{\partial
}{\partial t}\gg\frac{1}{n_{e0}}\frac{\partial\chi_{e0}}{\partial z}%
\frac{\partial}{\partial z}.
\end{equation}
The first inequality (57) means that the thermal conductivity is the dominant
mode of the thermal transport (Balbus 2000; Quataert 2008). Under the second
condition (57), we can neglect the inhomogeneity and perturbation of the
thermal flux conductivity coefficient in the temperature equation (50) (see 47
and 51). Obviously, the term proportional to $\partial^{2}\chi_{e0}/\partial
z^{2}$ in expression (51) can also be neglected. In this case, the small
correction proportional to $\partial/\partial t$ in the temperature equation
(50) which will be necessary for calculation of the electron velocity (see
below) will be larger than that $\sim\partial\chi_{e0}/\partial z$. We further
apply the operator $\partial/\partial t$ to (55) and use (44), (45), (49), and
(56). As a result, we obtain%
\begin{align}
\left(  c_{se}^{2}\frac{\partial}{\partial z}-\gamma g_{e}\right)
\frac{\partial v_{e1z}}{\partial z}  & \mathbf{=}\mathbf{-}\frac{\partial
F_{e1z}}{m_{i}\partial t}+\left[  \left(  1-\gamma\right)  g_{e}+c_{se}%
^{2}\frac{\partial}{\partial z}\right]  \frac{1}{m_{e}\omega_{ce}}%
\frac{\partial F_{e1x}}{\partial y}\\
& +\left(  \gamma-1\right)  \frac{\chi_{e0}}{m_{i}n_{e0}}\left(
\frac{\partial T_{e0}}{\partial z}\frac{1}{B_{0}}\frac{\partial^{2}B_{1y}%
}{\partial y\partial z}+\frac{\partial^{3}T_{e1}}{\partial z^{3}}\right)
.\nonumber
\end{align}
Equation for the temperature perturbation under conditions (57) has the form%

\begin{align}
\left[  \left(  \gamma-1\right)  \frac{\chi_{e0}}{n_{e0}}\frac{\partial^{2}%
}{\partial z^{2}}-\frac{\partial}{\partial t}\right]  T_{e1}  & =v_{e1z}%
\frac{\partial T_{e0}}{\partial z}+\left(  \gamma-1\right)  T_{e0}\left(
\frac{\partial v_{e1z}}{\partial z}-\frac{1}{m_{e}\omega_{ce}}\frac{\partial
F_{e1x}}{\partial y}\right) \\
& -\left(  \gamma-1\right)  \frac{\chi_{e0}}{n_{e0}}\frac{\partial T_{e0}%
}{\partial z}\frac{\partial B_{1y}}{B_{0}\partial y},\nonumber
\end{align}
where we have used (56).

To find equation for $v_{e1z}$, we substitute $T_{e1}$ from (59) into (58).
Taking into account a contribution of the term $\partial T_{e1}/\partial t$
and carrying out some transformations, we obtain%
\begin{align}
\frac{\partial^{3}v_{e1z}}{\partial z^{3}}  & =\mathbf{-}\frac{\partial
^{2}F_{e1z}}{T_{e0}\partial z\partial t}\mathbf{-}\frac{n_{e0}}{\chi_{e0}%
}\left(  \frac{\partial}{\partial z}\right)  ^{-1}\frac{\partial^{2}F_{e1z}%
}{T_{e0}\partial t^{2}}+\frac{1}{m_{e}\omega_{ce}}\frac{\partial^{3}F_{e1x}%
}{\partial y\partial z^{2}}\\
& +\frac{1}{c_{se}^{2}}\left(  \gamma g_{e}+\frac{\partial c_{se}^{2}%
}{\partial z}\right)  \frac{1}{m_{e}\omega_{ce}}\frac{\partial^{2}F_{e1x}%
}{\partial y\partial z}-\frac{\partial T_{e0}}{T_{e0}\partial z}\frac{1}%
{B_{0}}\frac{\partial^{2}B_{1y}}{\partial y\partial t}.\nonumber
\end{align}
The correction proportional to $\partial F_{e1x}/\partial t$ is absent. The
last term on the right hand-side of (60) is connected with the background
electron thermal flux.

From (59) and (60), we can find equation for the temperature perturbation
\begin{align}
\left(  \gamma-1\right)  \frac{\chi_{e0}}{n_{e0}}\frac{\partial}{\partial
z}\left(  \frac{\partial^{2}T_{e1}}{\partial z^{2}}+\frac{\partial T_{e0}%
}{\partial z}\frac{\partial B_{1y}}{B_{0}\partial y}\right)   & =\frac{\gamma
T_{e0}}{c_{se}^{2}}\left[  \left(  \gamma-1\right)  g_{e}+\frac{\partial
c_{se}^{2}}{\partial z}\right]  \frac{1}{m_{e}\omega_{ce}}\frac{\partial
F_{e1x}}{\partial y}\\
& -\left(  \gamma-1\right)  \frac{\partial F_{e1z}}{\partial t}-\gamma
\frac{n_{e0}}{\chi_{e0}}\left(  \frac{\partial}{\partial z}\right)  ^{-2}%
\frac{\partial^{2}F_{e1z}}{\partial t^{2}}\nonumber\\
& -\gamma\frac{\partial T_{e0}}{\partial z}\left(  \frac{\partial}{\partial
z}\right)  ^{-1}\frac{\partial^{2}B_{1y}}{B_{0}\partial y\partial t}.\nonumber
\end{align}
It is followed from results obtained below that all terms on the right-hand
side of (61) (except the correction $\sim\partial^{2}F_{e1z}/\partial t^{2} $)
have the same order of magnitude (see Sect. 4.3). The left-hand side of this
equation is larger (see conditions 57). Thus, the temperature perturbation in
the zero order of magnitude can be found by equaling the left part of (61) to
zero. However, the right part of this equation is necessary for finding the
transverse velocity perturbation $v_{e1x}$ (see below).

To find the velocity $v_{e1x}$, we need to calculate the value $P_{e1}$ (see
53 and 54). Performing calculations in the same way as that for ions (see
Sect. 4.3), we obtain%
\begin{align}
c_{se}^{2}\frac{\partial^{2}P_{e1}}{\partial z^{2}}  & =\left[  c_{se}%
^{2}\frac{\partial}{\partial z}+\left(  \gamma-1\right)  g_{e}+\frac{\partial
c_{se}^{2}}{\partial z}\right]  \left(  \mathbf{-}\frac{\partial F_{e1z}%
}{m_{i}\partial t}+\frac{\partial V_{e1}}{\partial z}\right) \\
& +g_{e}\left[  \left(  \gamma-1\right)  g_{e}+\frac{\partial c_{se}^{2}%
}{\partial z}\right]  \frac{1}{m_{e}\omega_{ce}}\frac{\partial F_{e1x}%
}{\partial y},\nonumber
\end{align}
where we have introduced notation connected with the thermal conductivity,%
\begin{equation}
V_{e1}=\left(  \gamma-1\right)  \frac{\chi_{e0}}{m_{i}n_{e0}}\left(
\frac{\partial T_{e0}}{\partial z}\frac{1}{B_{0}}\frac{\partial B_{1y}%
}{\partial y}+\frac{\partial^{2}T_{e1}}{\partial z^{2}}\right)  .
\end{equation}
Equation (62) can be re-written in the form which is convenient for finding
the velocity $v_{e1x}$. Using (61), we obtain%
\begin{align}
\frac{\partial^{2}}{\partial z^{2}}\left(  P_{e1}-V_{e1}\right)   &
=-\frac{\partial^{2}F_{e1z}}{m_{i}\partial z\partial t}-\frac{\gamma}%
{c_{se}^{2}}\left[  \left(  \gamma-1\right)  g_{e}+\frac{\partial c_{se}^{2}%
}{\partial z}\right]  \frac{\partial F_{e1z}}{m_{i}\partial t}\\
& +\frac{1}{c_{se}^{2}}\left[  \left(  \gamma-1\right)  g_{e}+\frac{\partial
c_{se}^{2}}{\partial z}\right]  \left(  \gamma g_{e}+\frac{\partial c_{se}%
^{2}}{\partial z}\right)  \frac{1}{m_{e}\omega_{ce}}\frac{\partial F_{e1x}%
}{\partial y}\nonumber\\
& -\left[  \left(  \gamma-1\right)  g_{e}+\frac{\partial c_{se}^{2}}{\partial
z}\right]  \frac{\partial T_{e0}}{T_{e0}\partial z}\left(  \frac{\partial
}{\partial z}\right)  ^{-1}\frac{\partial^{2}B_{1y}}{B_{0}\partial y\partial
t}.\nonumber
\end{align}

It is easy to see that (53) has the form%
\begin{equation}
\frac{\partial}{\partial t}\left(  v_{e1x}-\frac{1}{m_{e}\omega_{ce}}%
F_{e1y}\right)  \mathbf{=-}\frac{1}{\omega_{ci}}\frac{\partial}{\partial
y}\left(  P_{e1}-V_{e1}\right)  .
\end{equation}
Thus, the main contribution of the flux described by (63) does not influence
on the electron dynamics. Applying operator $\partial^{2}/\partial z^{2}$ to
(65) and using (64), we find an equation for the perturbed velocity $v_{e1x}$.

\bigskip

\section{FOURIER CURRENT COMPONENTS}

\subsection{Fourier velocity components of ions and electrons}

Let us give velocities of ions and electrons in the Fourier-representation.
From (20), (21), and (40), we have%
\begin{equation}
v_{i1xk}\mathbf{=}\frac{1}{\omega_{ci}^{2}}\left(  1+\frac{\omega^{2}}%
{\omega_{ci}^{2}}\right)  Q_{i1xk}+i\frac{k_{y}^{2}}{k_{z}^{2}}\frac{\left(
\omega^{2}-g_{i}a_{i}\right)  }{\omega\omega_{ci}^{3}}Q_{i1yk}\mathbf{-}%
\frac{1}{\omega_{ci}}\frac{k_{y}}{k_{z}}\left(  1-i\frac{a_{i}}{k_{z}}\right)
F_{i1zk},
\end{equation}%
\begin{equation}
v_{i1yk}\mathbf{=}\frac{1}{\omega_{ci}^{2}}\left[  1+\frac{\left(  k^{2}%
\omega^{2}-k_{y}^{2}g_{i}a_{i}\right)  }{k_{z}^{2}\omega_{ci}^{2}}\right]
Q_{i1yk}+i\frac{\omega}{\omega_{ci}^{2}}\frac{k_{y}}{k_{z}}\left(
1-i\frac{a_{i}}{k_{z}}\right)  F_{i1zk}.
\end{equation}
Here and below, we have introduced notations%
\begin{equation}
a_{i,e}=\frac{1}{c_{si,e}^{2}}\left[  \left(  \gamma-1\right)  g_{i,e}%
+\frac{\partial c_{si,e}^{2}}{\partial z}\right]  .
\end{equation}
The velocity $v_{i1zk}$ is given by (38).

For electrons, using (64) and (65), we find%
\begin{align}
v_{e1xk}  & \mathbf{=}\mathbf{-}i\frac{a_{e}c_{se}^{2}}{\omega\omega_{ci}%
}\frac{k_{y}^{2}}{k_{z}^{2}}\left(  b_{e}\frac{1}{m_{e}\omega_{ce}}%
F_{e1xk}+\omega\frac{\partial T_{e0}}{k_{z}T_{e0}\partial z}\frac{B_{1yk}%
}{B_{0}}\right) \\
& +\frac{1}{m_{e}\omega_{ce}}F_{e1yk}\mathbf{-}\frac{k_{y}}{k_{z}}\left(
1-i\gamma\frac{a_{e}}{k_{z}}\right)  \frac{1}{m_{e}\omega_{ce}}F_{e1zk}%
,\nonumber
\end{align}
where the following notation is introduced:
\begin{equation}
b_{e}=\frac{1}{c_{se}^{2}}\left(  \gamma g_{e}+\frac{\partial c_{se}^{2}%
}{\partial z}\right)  .
\end{equation}
Equation (60) gives us
\begin{align}
v_{e1zk}  & =\frac{k_{y}}{k_{z}}\frac{1}{m_{e}\omega_{ce}}F_{e1xk}%
-i\frac{k_{y}}{k_{z}^{2}}\left(  b_{e}\frac{1}{m_{e}\omega_{ce}}%
F_{e1xk}+\omega\frac{\partial T_{e0}}{k_{z}T_{e0}\partial z}\frac{B_{1yk}%
}{B_{0}}\right) \\
& -i\frac{\omega}{k_{z}^{2}T_{e0}}\left(  1+i\omega\frac{n_{e0}}{\chi
_{e0}k_{z}^{2}}\right)  F_{e1zk}.\nonumber
\end{align}
The velocity $v_{e1y}$ is defined by (52).

\subsection{Fourier electron velocity components at the absence of the heat
flux}

To elucidate the role of the electron thermal flux, we will also consider the
dispersion relation when this flux is absent. Therefore, we also give here the
corresponding electron velocity components:
\begin{equation}
v_{e1xk}=-i\frac{k_{y}^{2}g_{e}a_{e}}{k_{z}^{2}\omega\omega_{ci}}\frac
{1}{m_{e}\omega_{ce}}F_{e1xk}+\frac{1}{m_{e}\omega_{ce}}F_{e1yk}-\frac{k_{y}%
}{k_{z}}\left(  1-i\frac{a_{e}}{k_{z}}\right)  \frac{1}{m_{e}\omega_{ce}%
}F_{e1zk},
\end{equation}%
\begin{equation}
v_{e1zk}\mathbf{=}\frac{k_{y}}{k_{z}}\left(  1-i\frac{g_{e}}{k_{z}c_{se}^{2}%
}\right)  \frac{1}{m_{e}\omega_{ce}}F_{e1xk}\mathbf{-}i\frac{\omega}{k_{z}%
^{2}c_{se}^{2}m_{i}}\left(  1-i\frac{\gamma g_{e}}{k_{z}c_{se}^{2}}\right)
F_{e1zk}.
\end{equation}
Comparing expressions (69) and (71) with these equations, we see that the
thermal flux essentially modifies the small terms in the electron velocity
under conditions (57).

\subsection{Fourier components of current}

We find now the Fourier components of the linear current $\mathbf{j}_{1}%
=q_{i}n_{i0}\mathbf{v}_{i1}+q_{e}n_{e0}\mathbf{v}_{e1}$. It is convenient to
consider the value $4\pi i\mathbf{j}_{1}/\omega$. Using expressions (38),
(52), (66), (67), (69), and (71), we obtain the following current components:%
\begin{align}
\frac{4\pi i}{\omega}j_{1xk}  & =a_{xx}E_{1xk}+ia_{xy}E_{1yk}-a_{xz}E_{1zk}\\
& -b_{xx}\left(  v_{i1xk}-v_{e1xk}\right)  -ib_{xy}\left(  v_{i1yk}%
-v_{e1yk}\right)  +b_{xz}\left(  v_{i1zk}-v_{e1zk}\right)  ,\nonumber
\end{align}
\ \
\begin{align}
\frac{4\pi i}{\omega}j_{1yk}  & =-ia_{yx}E_{1xk}+a_{yy}E_{1yk}-a_{yz}E_{1zk}\\
& +ib_{yx}\left(  v_{i1xk}-v_{e1xk}\right)  -b_{yy}\left(  v_{i1yk}%
-v_{e1yk}\right)  +b_{yz}\left(  v_{i1zk}-v_{e1zk}\right)  ,\nonumber
\end{align}%
\begin{align}
\frac{4\pi i}{\omega}j_{1zk}  & =-a_{zx}E_{1xk}-a_{zy}E_{1yk}+a_{zz}E_{1z}\\
& +b_{zx}\left(  v_{i1x}-v_{e1x}\right)  +b_{zy}\left(  v_{i1y}-v_{e1y}%
\right)  -b_{zz}\left(  v_{i1z}-v_{e1z}\right)  .\nonumber
\end{align}
When obtaining (74)-(76), we have used notations (16), (23), (24), and (48)
and equalities $q_{e}=-q_{i}$, $n_{e0}=n_{i0}$, $m_{e}\nu_{ei}=m_{i}\nu_{ie}$.
We also have substituted $B_{1yk}$ by $(k_{z}c/\omega)E_{1xk}$, using (9). The
following notations are introduced in (74)-(76):%
\begin{align}
a_{xx}  & =\frac{\omega_{pi}^{2}}{\omega_{ci}^{2}}\frac{k^{2}}{k_{z}^{2}%
}\left(  1-\frac{k_{y}^{2}}{k^{2}}\frac{g_{i}a_{i}+a_{e}b_{e}c_{se}^{2}%
}{\omega^{2}}-\frac{k_{y}^{2}}{k^{2}}\frac{a_{e}c_{se}^{2}}{\omega^{2}}%
\frac{\partial T_{e0}^{\ast}}{T_{e0}\partial z}\right)  ,\\
a_{xy}  & =a_{yx}=\frac{\omega_{pi}^{2}\omega}{\omega_{ci}^{3}}\frac{k^{2}%
}{k_{z}^{2}}\left(  1-\frac{k_{y}^{2}}{k^{2}}\frac{g_{i}a_{i}}{\omega^{2}%
}\right)  ,a_{xz}=\frac{\omega_{pi}^{2}}{\omega\omega_{ci}}\frac{k_{y}}%
{k_{z}^{2}}\left(  a_{i}-\gamma a_{e}\right)  ,\nonumber\\
a_{yy}  & =\frac{\omega_{pi}^{2}}{\omega_{ci}^{2}},a_{yz}=a_{zy}=\frac
{\omega_{pi}^{2}}{\omega_{ci}^{2}}\frac{k_{y}}{k_{z}},a_{zx}=\frac{\omega
_{pi}^{2}}{\omega\omega_{ci}}\frac{k_{y}}{k_{z}^{2}}\left(  b_{e}-\frac{g_{i}%
}{c_{si}^{2}}+\frac{\partial T_{e0}^{\ast}}{T_{e0}\partial z}\right)
,\nonumber\\
a_{zz}  & =\frac{\omega_{pi}^{2}}{k_{z}^{2}}\left(  \frac{\gamma}{c_{se}^{2}%
}+\frac{1}{c_{si}^{2}}\right) \nonumber
\end{align}
and%
\begin{align}
b_{xx}  & =\frac{\omega_{pi}^{2}\nu_{ie}}{\omega_{ci}^{2}}\frac{m_{i}}{q_{i}%
}\frac{k^{2}}{k_{z}^{2}}\left(  1-\frac{k_{y}^{2}}{k^{2}}\frac{g_{i}%
a_{i}+a_{e}c_{se}^{2}b_{e}}{\omega^{2}}\right)  ,\\
b_{zx}  & =\frac{\omega_{pi}^{2}}{\omega\omega_{ci}}\frac{k_{y}}{k_{z}^{2}%
}\left(  b_{e}-\frac{g_{i}}{c_{si}^{2}}\right)  \frac{m_{i}}{q_{i}}\nu
_{ie},\nonumber\\
b_{ij}  & =a_{ij}\frac{m_{i}}{q_{i}}\nu_{ie}.\nonumber
\end{align}
Here $\omega_{pi}=\left(  4\pi n_{i0}q_{i}^{2}/m_{i}\right)  ^{1/2}$ is the
ion plasma frequency and $k^{2}=k_{y}^{2}+k_{z}^{2}$. The terms proportional
to $T_{e0}^{\ast}$ are connected with the background electron thermal flux.

Calculations show that to obtain expressions for $a_{ij}$ without thermal
flux, using electron velocities (72) and (73), we must change $b_{e}$ by
$g_{e}/c_{se}^{2}$, put $T_{e0}^{\ast}=0$, and take $\gamma=1$ in terms
$a_{xz}$ and $a_{zz}$.

\subsection{Simplification of collision contribution}

An assumption that electrons are magnetized has only been involved by
neglecting the transverse electron thermal flux. In other respects, a
relationship between $\omega_{ce}$ and $\nu_{ei}$ or $\omega_{ci}$ and
$\nu_{ie}$ (that is the same) can be arbitrary in (74)-(76). We further
proceed by taking into account that $\omega\ll\omega_{ci}$. In this case, we
can neglect the collisional terms proportional to $b_{xy}$ and $b_{yx}$.
However, the system of equations (74)-(76) stays sufficiently complex to find
$\mathbf{j}_{1}$ through $\mathbf{E}_{1}$. Therefore, we further consider the
case in which the frequency $\omega$ and wave numbers satisfy the following
conditions:%
\begin{equation}
\frac{\omega_{ci}^{2}}{\nu_{ie}^{2}}\frac{k_{z}^{2}}{k^{2}}\gg\frac{\omega
}{\nu_{ie}}\gg\frac{1}{k_{z}^{2}H^{2}}\frac{k_{y}^{2}c_{s}^{2}}{\omega
_{ci}^{2}},
\end{equation}
where%
\begin{equation}
c_{s}^{2}=\frac{c_{si}^{2}c_{se}^{2}}{\gamma c_{si}^{2}+c_{se}^{2}}.
\end{equation}
It is clear that conditions (79) can easily be realized. In this case, the
current components are equal to%
\begin{align}
\frac{4\pi i}{\omega}j_{1xk}  & =\varepsilon_{xx}E_{1xk}+i\varepsilon
_{xy}E_{1yk}-\varepsilon_{xz}E_{1zk},\\
\frac{4\pi i}{\omega}j_{1yk}  & =-i\varepsilon_{yx}E_{1xk}+\varepsilon
_{yy}E_{1yk}-\varepsilon_{yz}E_{1zk},\nonumber\\
\frac{4\pi i}{\omega}j_{1zk}  & =-\varepsilon_{zx}E_{1xk}-\varepsilon
_{zy}E_{1yk}+\varepsilon_{zz}E_{1z}.\nonumber
\end{align}
Components of the dielectric permeability tensor $\varepsilon_{ij}$ are the
following:
\begin{align}
\varepsilon_{xx}  & =a_{xx}+i\frac{\nu_{ie}}{\omega_{ci}}\frac{k_{y}}%
{k_{z}^{2}}\frac{\left(  a_{i}-\gamma a_{e}\right)  }{\left(  1-id_{z}\right)
}a_{zx},\varepsilon_{xy}=a_{xy}+\frac{\nu_{ie}}{\omega_{ci}}\frac{k_{y}}%
{k_{z}^{2}}\frac{\left(  a_{i}-\gamma a_{e}\right)  }{\left(  1-id_{z}\right)
}a_{zy},\\
\varepsilon_{xz}  & =\frac{a_{xz}}{\left(  1-id_{z}\right)  },\varepsilon
_{yx}=a_{yx}-\frac{\omega\nu_{ie}}{\omega_{ci}^{2}}\frac{k_{y}}{k_{z}}%
\frac{a_{zx}}{\left(  1-id_{z}\right)  },\varepsilon_{yy}=a_{yy},\nonumber\\
\varepsilon_{yz}  & =\frac{a_{yz}}{\left(  1-id_{z}\right)  },\varepsilon
_{zx}=\frac{a_{zx}}{\left(  1-id_{z}\right)  },\varepsilon_{zy}=\frac{a_{zy}%
}{\left(  1-id_{z}\right)  },\varepsilon_{zz}=\frac{a_{zz}}{\left(
1-id_{z}\right)  },\nonumber
\end{align}
where we have used notations (78). Parameter $d_{z}$,
\begin{equation}
d_{z}=\frac{\omega\nu_{ie}}{k_{z}^{2}c_{s}^{2}},
\end{equation}
defines the collisionless, $d_{z}\ll1$, and collisional, $d_{z}\gg1$, regimes.
Below, we derive the dispersion relation.

\bigskip

\section{DISPERSION\ RELATION}

Using (9) and (10) in the Fourier-representation and a system of equations
(81), we obtain the following equations for the electric field components:
\begin{align}
\left(  n^{2}-\varepsilon_{xx}\right)  E_{1xk}-i\varepsilon_{xy}%
E_{1yk}+\varepsilon_{xz}E_{1zk}  & =0,\\
i\varepsilon_{yx}E_{1xk}+\left(  n_{z}^{2}-\varepsilon_{yy}\right)
E_{1yk}+\left(  -n_{y}n_{z}+\varepsilon_{yz}\right)  E_{1zk}  & =0,\nonumber\\
\varepsilon_{zx}E_{1xk}+\left(  -n_{y}n_{z}+\varepsilon_{zy}\right)
E_{1yk}+\left(  n_{y}^{2}-\varepsilon_{zz}\right)  E_{1zk}  & =0,\nonumber
\end{align}
where $\mathbf{n=k}c/\omega$. The dispersion relation can be found by setting
the determinant of the system (84) equal to zero. In our case, the terms
proportional to $\varepsilon_{xy}$ and $\varepsilon_{yx}$ can be neglected. As
a result, we have
\[
\left(  n^{2}-\varepsilon_{xx}\right)  \left[  n_{y}^{2}\varepsilon
_{yy}+\left(  n_{z}^{2}-\varepsilon_{yy}\right)  \varepsilon_{zz}-n_{y}%
n_{z}\left(  \varepsilon_{yz}+\varepsilon_{zy}\right)  +\varepsilon
_{yz}\varepsilon_{zy}\right]
\]%
\begin{equation}
+\left(  n_{z}^{2}-\varepsilon_{yy}\right)  \varepsilon_{xz}\varepsilon
_{zx}=0.
\end{equation}
This dispersion relation can be studied for different cases. In subsequent
sections, we consider both the collisionless and collisional cases.

\subsection{ Collisionless case}

We assume that condition%
\begin{equation}
\frac{\omega\nu_{ie}}{k_{z}^{2}c_{s}^{2}}\ll1,
\end{equation}
is satisfied. Then, using notations (77) and (82), we reduce the dispersion
relation (85) to the form%
\begin{equation}
\left(  \omega^{2}-k_{z}^{2}c_{A}^{2}\right)  \left(  \omega^{2}-k_{z}%
^{2}c_{A}^{2}-\Omega^{2}\frac{k_{y}^{2}}{k^{2}}\right)  =0,
\end{equation}
where $c_{A}=B_{0}/(4\pi m_{i}n_{i0})^{1/2}$ is the Alfv\'{e}n velocity and
\begin{equation}
\Omega^{2}=g_{i}a_{i}+c_{se}^{2}a_{e}b_{e}+c_{se}^{2}a_{e}\frac{\partial
T_{e0}^{\ast}}{T_{e0}\partial z}+c_{s}^{2}\left(  a_{i}-\gamma a_{e}\right)
\left(  b_{e}-\frac{g_{i}}{c_{si}^{2}}+\frac{\partial T_{e0}^{\ast}}%
{T_{e0}\partial z}\right)  .
\end{equation}
When obtaining (87), we have used the condition $k_{y}^{2}c_{s}^{2}%
/\omega_{ci}^{2}\ll1$. We see that there are two wave modes. The first mode,
$\omega^{2}=k_{z}^{2}c_{A}^{2}$, is the Alfv\'{e}n wave with a polarization of
the electric field mainly along the $y$-axis (remind that the wave vector
$\mathbf{k}$ is situated in the $y-z$ plane). This wave does not feel the
inhomogeneity of medium. The second wave has a polarization of the
magnetosonic wave, i.e. its electric field is directed mainly along the
$x$-axis (see below). This wave is undergone by the action of the medium
inhomogeneity effect. The corresponding dispersion relation is
\begin{equation}
\omega^{2}=k_{z}^{2}c_{A}^{2}+\Omega^{2}\frac{k_{y}^{2}}{k^{2}}.
\end{equation}
Expression (88) can further be simplified, using (11), (12), (68), (70), and
(80). As a result, we obtain%
\begin{equation}
\Omega^{2}=\frac{\gamma}{\left(  \gamma c_{si}^{2}+c_{se}^{2}\right)
m_{i}^{2}}\left[  \left(  \gamma-1\right)  m_{i}g+\gamma\frac{\partial\left(
T_{i0}+T_{e0}\right)  }{\partial z}\right]  \left[  m_{i}g+\frac
{\partial\left(  T_{e0}+T_{e0}^{\ast}\right)  }{\partial z}\right]  .
\end{equation}

We have pointed out at the end of Sect. 6.3 what changes must be done in
expressions (77) and (78) to consider the case without the electron heat flux.
This case follows from (90), if we omit the term $\partial\left(
T_{e0}+T_{e0}^{\ast}\right)  /\partial z$ and put $\gamma=1$ in the first
multiplier. Then $\Omega^{2}$ becomes ($\Omega^{2}\rightarrow\Omega_{1}^{2}$)%

\begin{equation}
\Omega_{1}^{2}=\frac{g}{\left(  c_{si}^{2}+c_{se}^{2}\right)  }\left[  \left(
\gamma-1\right)  g+\frac{\partial\left(  c_{si}^{2}+c_{se}^{2}\right)
}{\partial z}\right]  .
\end{equation}
This is the Brunt-V\"{a}is\"{a}l\"{a} frequency. Comparing expressions (90)
and (91), we see that the heat flux stabilizes the unstable ($\Omega_{1}%
^{2}<0$) stratification. We also see from (90) that the background heat flux
($\sim T_{e0}^{\ast}$) has no a fundamental importance. If the temperature
decreases in the direction of gravity ($\partial T_{i,e0}/\partial z>0$), the
medium is stable. Solution (90) describes an instability only when
\[
\frac{\gamma-1}{2\gamma}m_{i}g<-\frac{\partial T_{0}}{\partial z}<\frac{1}%
{2}m_{i}g,
\]
where $T_{i0}\sim T_{e0}=T_{0}$. We also note that $\Omega^{2}$ can be
negative if gradients of $T_{i0}$ and $T_{e0}$ have different signs.

The dispersion relation (87) with $\Omega^{2}$ defined by (90) considerably
differs from the dispersion relation obtained in the framework of the ideal
MHD (Quataert 2008). The reasons of this are discussed in Sect. 8.

\subsection{Collisional case}

We proceed with the collisional case when
\begin{equation}
\frac{\omega\nu_{ie}}{k_{z}^{2}c_{s}^{2}}\gg1.
\end{equation}
In this limiting case, the dispersion relation takes the form $n^{2}%
-\varepsilon_{xx}\approx0$ and we obtain again (89). Thus, the dispersion
relation is the same for both the collisionless and collisional cases. We note
that this result has also been obtained for the case in which gravity is
perpendicular to the magnetic field (Nekrasov and Shadmehri 2010).

\bigskip

\subsection{Polarization of perturbations}

Let us neglect in the system of equations (84) the small contributions given
by $\varepsilon_{xy}$ and $\varepsilon_{yx}$. Then, for example, in the
collisionless case, we obtain for the second wave $\omega^{2}\neq k_{z}%
^{2}c_{A}^{2}$,%
\begin{align}
E_{1yk}  & =\frac{k_{y}}{k_{z}}E_{1zk},\\
E_{1zk}  & =\frac{\varepsilon_{zx}}{\varepsilon_{zz}}E_{1xk}\ll E_{1xk}%
.\nonumber
\end{align}
Thus, the second wave has a polarization of the electric field mainly along
the $x$-axis. In spite of that the component $E_{1zk}\ll E_{1xk}$, it is
multiplied by a large coefficient in the first equation of the system (84). As
a result, the contribution of this term is the same on the order of magnitude
as that of the first term.

In the collisional case, the component $E_{1zk}$\ is also defined by (93).
However, its contribution to the first equation of the system (84) can be
neglected. However, the contribution of the collisional term connected with
the longitudinal current in (74) which is proportional to $E_{1xk}$ in (76) is important.

\bigskip

\section{DISCUSSION}

In the MHD analysis of the buoyancy instabilities, one assumes that
$p_{i1k}/p_{i0}$ $\ll n_{i1k}/n_{i0}$ (see, e.g., Balbus 2000; Quataert 2008;
Ren et al. 2009). This relation is correct for internal gravity waves in the
neutral medium (e.g., Nekrasov 1994). It is also correct for perturbations
$n_{i1k}$ and $p_{i1k}$ connected with the transverse perturbations $Q_{i1yk}$%
\[
\frac{p_{i1k}}{p_{i0}}/\frac{n_{i1k}}{n_{i0}}\left(  \sim Q_{i1yk}\right)
\sim\frac{g_{i}}{k_{z}c_{si}^{2}}\ll1
\]
(see 41 and 42). However, it is followed from the last equations that due to
the longitudinal electric field perturbation $E_{1zk}$ which is of the order
of $E_{1zk}\sim\left(  k_{y}c_{s}^{2}/\omega\omega_{ci}H\right)  E_{1xk}$ (see
77, 82, and 93) the relative pressure and number density perturbations are of
the same order of magnitude
\[
\frac{p_{i1k}}{p_{i0}}/\frac{n_{i1k}}{n_{i0}}\left(  \sim E_{1zk}\right)
\sim1.
\]
The ideal MHD does not capture the field $E_{1z}$. Therefore, results obtained
in the MHD framework and multicomponent plasma approach are different. In
Sect. 4.3, we also have shown that $\mathbf{\nabla\cdot v}_{i1}\sim
(g_{i}/c_{si}^{2})v_{i1z}$. This relation is also true for internal gravity
waves (Nekrasov 1994). The fact that $\mathbf{\nabla\cdot v}_{i1}\neq0$ for
gaseous media is taken into account when deriving an internal energy equation
($\mathbf{\nabla\cdot v}_{i1}$ is excluded from the number density and
pressure or temperature equations). Then in the MHD framework, one can use the
condition of incompressibility $\mathbf{\nabla\cdot v}_{i1}=0$ in the momentum
and magnetic induction equations for perturbations much slower than the sound
waves. In our case, the divergence of the velocity defined by the main terms
in $\mathbf{v}_{i1k}\sim Q_{i1yk}$ (see 38 and 67) and $\mathbf{v}_{e1k}\sim
F_{e1xk}$ (see 52 and 71) is also equal to zero. However, these main terms are
the same for ions and electrons and canceled with each other at calculation of
the current. Therefore, together with velocity perturbations proportional to
the longitudinal force $F_{i,e1zk}$, we must take into account contribution of
additional small velocities connected with transverse perturbations
$Q_{i1x,yk}$ and $F_{e1xk} $.

From the dispersion relation (85), we see the necessity of involving the
contribution of values $\varepsilon_{xz}$, $\varepsilon_{zx}$, and
$\varepsilon_{zz}$ in\textbf{\ }the collisionless case (86) (values
$\varepsilon_{xz}$ and $\varepsilon_{zx}$ give the last term on the right
hand-side of 88). This means that contribution of currents $j_{1x}\sim E_{1z}
$ and $j_{1z}\sim E_{1x},E_{1z}$ must be taken into account. As for the
collisional case (92), the electric field $E_{1z}$ is not important. In the
current $j_{1xk}$, we must take into consideration the contribution of the
current $j_{1zk}$\ as a result of collisions which is proportional to
$E_{1xk}$\ (see 74 and 76). This collisional case also is not captured by the
ideal MHD.

Thus, the standard MHD equations with simplified assumptions are not
applicable for the correct theory of buoyancy instabilities. Such a theory can
only be given by the multicomponent approach used in this paper.

The results following from (90) show that the thermal flux stabilizes the
buoyancy instability in the case of the geometry under consideration. The
instability only is possible in the narrow region of the temperature gradient
(see Sect. 7.1). The presence of the background electron thermal flux (the
term $\sim T_{e0}^{\ast}$) does not play an essential role. An instability
also is possible, if the temperature gradients of ions and electrons have the
opposite signs.

The contribution of collisions between electrons and ions depends on the
parameter $d_{z}$ defined by (83). In the both limits (86) ($d_{z}\ll1$) and
(92) ($d_{z}\gg1$), the dispersion relation has the same form.

In our analysis, we have considered for generality that ions and electrons
have different temperatures. However in (3), (6), and (7), the terms
describing the energy exchange between species due to their collisions has not
been taken into account. This is possible, if the dynamical timescale is
smaller than the timescale of smoothing of the ion and electron temperatures,
i.e. $\nu_{ie}\ll\Omega$. In the opposite case, $\nu_{ie}\gg\Omega$, the
perturbed temperatures of electrons and ions are almost equal one another.
Equations (6) and (7) for electrons will keep their form because
$\mathbf{v}_{e1}\approx\mathbf{v}_{i1}$. In the case $T_{e0}\approx T_{i0}$,
these equations will stay the same with the heat flux two times less than the
former one. \

Conditions of our consideration (33) and (35) are satisfied when $1\gg\rho
_{i}/H$ and $1\gg k_{z}Hk_{y}^{2}\rho_{i}^{2}$, where $\rho_{i}$ is the ion
Larmor radius. For estimations, we take $\omega\sim g/c_{s}$ $(T_{e0}\sim
T_{i0})$. It is obvious that these inequalities can be justified. It is easy
to verify that conditions (79) are also satisfied. However, conditions (57)
can impose some restrictions. In the case $T_{e0}\sim T_{i0}$, they can be
written in the form
\begin{equation}
1\gg\frac{k_{z}c_{s}}{\nu_{ie}}\gg\frac{1}{k_{z}H},
\end{equation}
where we have used expression for $\chi_{e0}$ (Braginskii 1965). From
inequalities (94), it is in particular followed that the case of consideration
is justified when $\nu_{ie}\gg\omega$ and $d_{z}\ll1$. In the case $T_{e0}\gg
T_{i0}$, the value $d_{z}$ is in the limits%
\[
\frac{T_{e0}}{T_{i0}}\gg d_{z}\gg\frac{T_{e0}}{T_{i0}}\frac{1}{k_{z}H}
\]
and can be both $<1$ and $\gtrsim1$.

We will further compare results obtained in this paper with the case
when stratification is perpendicular to the magnetic field (Nekrasov and
Shadmehri, 2010). However, first of all, we would like to say a few words
about the Schwarzschild criterion of the buoyancy instability. It is generally
accepted that this instability is possible, if the entropy increases in the
direction of gravity. From a formal point of view, it is correct, if we take
the Brunt-V\"{a}is\"{a}l\"{a} frequency $N$ in the form (e.g. Balbus
2000),%
\[
N^{2}=-\frac{1}{\gamma\rho}\frac{\partial p}{\partial z}\frac{\partial\ln
p\rho^{-\gamma}}{\partial z}.
\]
However, this expression can easily be transformed into expression (32).
Thus, we see that the buoyancy instability exists, if the temperature
increases along the gravity and the temperature gradient exceeds a certain
threshold.

When the thermal conduction is the dominant process in the electron
temperature evolution, the buoyancy instability in the case $g\perp
B_{0}$ can arise according to criterion which is analogous to the
Schwarzschild criterion (see 65 in Nekrasov and Shadmehri, 2010). The same is
also true when the thermal conduction is negligible (66 in Nekrasov and
Shadmehri, 2010). Both criteria are similar. Thus, we can conclude that from the
point of view of observations it is difficult to define the role of thermal
conduction in generation of instability and turbulence. However, in the case
$g\parallel B_{0}$, the situation is different. For the
negligible electron heat flux, we have the similar criterion of instability as
that for $g\perp B_{0}$(91). When the thermal conduction
is dominant, the possible instability at $\partial\left(
T_{i0}+T_{e0}\right)  /\partial z<0$\ is stabilized (90). These
results could be used by observers for determination of the mutual orientation
of the magnetic field and gravity in astrophysical objects, e.g., in
galaxy clusters. 

The true geometry of the magnetic field lines in the ICM is poorly
understood. However, one may consider two extreme cases for the direction of
the dominant magnetic field lines depending on the direction of gravity.
The direction of the latter can have a vital role in driving turbulence via the
possible effect of convective heat flux. Since this flux is mainly along the
magnetic field lines, the two extreme cases are considered as either the
magnetic field is perpendicular to the gravity or parallel to it. Thus, the
true response of ICM to small perturbations would be possibly between
these two cases. Of course, when the system evolves and enters into the
nonlinear regime, one may expect saturation of the instability by rearranging
the magnetic field lines. Measurements of the magnetic field in ICM are not
consistent and there is a factor of four to ten of discrepancy depending on
the method (e.g., Carilli and Taylor 2002). Physical mechanisms that may
affect these observational measurements of the magnetic field in ICM are very
important in this regard. Our analysis gives a better understanding of
such a mechanism, i.e. buoyancy instability, though more detailed work is needed in future.

\textbf{\bigskip}

\section{\bigskip CONCLUSION}

In this paper, we have investigated buoyancy instabilities in magnetized
electron-ion plasmas with the anisotropic electron thermal flux, using the
multicomponent approach when the dynamical equations for the ions and
electrons are solved separately via electric field perturbations. We have
included the background electron heat flux and collisions between electrons
and ions. The important role of the longitudinal electric field perturbation,
which is not captured by the ideal MHD equations, has been demonstrated. We
have shown that the previous MHD\ result for the growth rate in the geometry
considered in this paper when all background quantities are directed along the
one axis is questionable. The reason of this has been shown to be in
simplified assumptions made in the MHD analysis of the buoyancy instabilities
and some shortcomings of the MHD.

At the consideration of the electron heat flux, we have adopted that the
electron cyclotron frequency is much larger than the electron collision
frequency that is typical for tokamaks, solar corona, and astrophysical
objects such as ICM and galaxy clusters. The dispersion relation obtained
shows that the anisotropic electron heat flux including the background one
stabilizes the unstable stratification except the narrow region of the
temperature gradient. However, when gradients of the ion and electron
temperatures have opposite signs, the medium becomes unstable.

Results obtained in this paper are applicable to the magnetized stratified
objects and can be useful for searching sources of turbulent transport of
energy and matter. For astrophysical plasmas, it has been suggested that the
buoyancy instability can act as a driving mechanism to generate turbulence in
ICM and this extra source of heating may help to resolve the cooling flow
problem (e.g., Allen 2000). However, all previous analytical and numerical
studies are restricted to the MHD approach. Our study shows that when the true
multifluid nature of the system with the electron heat flux is considered, one
can not expect the buoyancy instability unless for a very limited range of the
gradient of the temperature or when the gradients of the temperature of the
electrons and ions have opposite signs. Both cases are very unlikely. However,
in the case when the heat flux does not play the role, the system can be
unstable due to the convective instability.

\subsection{Acknowledgments}

We gratefully thanks the anonymous referee for his/her constructive and useful
comments which have helped to improve this paper.

\bigskip\

\subsection{ References}

\bigskip

Allen, S.W.: Mon. Not. R. Astron. Soc. \textbf{315}, 269 (2000).

Balbus, S.A.: Astrophys. J.\textbf{\ 534}, 420 (2000).

Balbus, S.A.: Astrophys. J. \textbf{562}, 909 (2001).

Betti, R., Goncharov, V.N., McCrory, R.L., Verdon, C.P.: Phys. Plasmas
\textbf{5}, 1446 (1998).

Bian, N.H., Kontar, E.P.: Phys. Plasmas \textbf{17}, 062308 (2010).

Braginskii, S.I.: Rev. Plasma Phys. \textbf{1}, p. 205 (1965).

Canaud, B., Fortin, X., Garaude, F., Meyer, C., Philippe, F., Temporal, M.,
Atzeni, S., Schiavi, A.: Nucl. Fusion \textbf{44, }1118 (2004).

Carilli, C.L., Taylor, G.B.: Annu. Rev. Astron. Astrophys. \textbf{40}, 319 (2002).

Chang, P., Quataert, E.: Mon. Not. R. Astron. Soc. \textbf{403}, 246 (2010).

Dimits, A.M., Bateman, G., Beer, M.A., Cohen, B.I., Dorland, W., Hammett,
G.W., Kim, C., Kinsey, J.E., Kotschenreuther, M., Kritz, A.H., Lao, L.L.,
Mandrekas, J., Nevins, W.M., Parker, S.E., Redd, A.J., Shumaker, D.E., Sydora,
R., Weiland, J.: Phys. Plasmas \textbf{7}, 969 (2000).

Fabian, A.C., Sanders, J.S., Taylor, G.B., Allen, S.W., Crawford, C.S.,
Johnstone, R.M., Iwasawa, K.: Mon. Not. R. Astron. Soc. \textbf{366}, 417 (2006).

Fryxell, B., Mueller, E., Arnett, D.: Astrophys. J. \textbf{367}, 619 (1991).

Garbet, X.: Plasma Phys. Controlled Fusion \textbf{43}, A251 (2001).

Gossard, E.E., Hooke, W.H.: Waves in the Atmosphere, Elsevier Scientific
Publishing Company, Amsterdam (1975).

Isobe, H., Miyagoshi, T., Shibata, K., Yokoyama, T.: Nature (London),
\textbf{434}, 478 (2005).

Kadomtsev, B.B., Pogutse, O.P.: In: Leontovich, M.A. (ed.) Review of Plasma
Physics\textit{, } \textit{\ }Consultants Bureau, New York, \textbf{5}, p. 249
(1965).\textbf{\ }

Kojima, Y., Oogi, J.: Mon. Not. R. Astron. Soc. \textbf{398}, 271 (2009).

Lindl, J.D., Amendt, P., Berger, R.L., Glendinning, S.G., Glenzer, S.H., Haan,
S.W., Kauffman, R.L., Landen, O.L., Suter, L.J.: Phys. Plasmas \textbf{11},
339 (2004).

McClements, K.G., Fletcher, L.: Astrophys. J. \textbf{693}, 1494 (2009).

Narayan, R., Igumenshchev, I.V., Abramowicz, M.A.: Astrophys. J. \textbf{539},
798 (2000).

Narayan, R., Quataert, E., Igumenshchev, I.V., Abramowicz, M.A.: Astrophys. J.
\textbf{577}, 295 (2002).

Nekrasov, A.K.: J. Atmos. Terr. Phys. \textbf{56}, 931 (1994).

Nekrasov, A.K.: Phys. Plasmas \textbf{15}, 032907 (2008).

Nekrasov, A.K.: Phys. Plasmas \textbf{16}, 032902 (2009a).

Nekrasov, A.K.: Astrophys. J. \textbf{695}, 46 (2009b).

Nekrasov, A.K.: Astrophys. J. \textbf{704}, 80 (2009c).

Nekrasov, A.K.: Mon. Not. R. Astron. Soc. \textbf{400}, 1574 (2009d).

Nekrasov, A.K., Shadmehri, M.: Astrophys. J. \textbf{724, }1165 (2010).

Parrish, I.J., Quataert, E., Sharma, P.: Astrophys. J. \textbf{703}, 96 (2009).

Pedlosky, J.: Geophysical Fluid Dynamics, Springer-Verlag, New York (1982).

Quataert, E.: Astrophys. J. \textbf{673}, 758 (2008).

Ren, H., Wu, Z., Cao, J., Chu, P.K.: Phys. Plasmas \textbf{16}, 102109 (2009).

Ren, H., Wu, Z., Cao, J., Chu, P.K., Li, D.: Phys. Plasmas \textbf{17}, 042117 (2010a).

Ren, H., Wu, Z., Dong, C., Chu, P.K.: Phys. Plasmas \textbf{17}, 052102 (2010b).

Sanders, J.S., Fabian, A.C., Frank, K.A., Peterson, J.R., Russell, H.R.: Mon.
Not. R. Astron. Soc. \textbf{402}, 127 (2010).

Sarazin, C.L.: X-Ray Emission from Clusters of Galaxies, Cambridge Univ.
Press, Cambridge (1988).

Schwarzschild, M.: Structure and Evolution of the Stars, Dover, New York (1958).

Spitzer, L., Jr.: Physics of Fully Ionized Gases, 2d ed., Interscience, New
York (1962).

Takahashi, S., Kawai, H., Ohsawa, Y., Usami, S., Chiu, C., Horton, W.: Phys.
Plasmas \textbf{16}, 112308 (2009).

\bigskip

\bigskip

\bigskip
\end{document}